\documentclass[pre,aps,reprint,amsmath]{revtex4-1}

\usepackage[dvips]{graphicx}
\usepackage{amssymb,amsfonts,amsmath}
\usepackage{color}
\usepackage{ulem}
\usepackage[hidelinks]{hyperref}
\usepackage{bbm}
\usepackage{upgreek}

\newcommand{\rv}{{\mathbf r}}

\newcommand{\Tr}{{\rm Tr}\,}

\newcommand{\pv}{{\bf p}}

\newcommand{\Fv}{{\bf F}}

\newcommand{\eps}{{\boldsymbol \epsilon}}
\newcommand{\unity}{{\mathbbm 1}}
\newcommand{\tot}{{\rm o}}
\newcommand{\ext}{{\rm ext}}
\newcommand{\exc}{{\rm exc}}

\begin{document}

\title{Variance of fluctuations from Noether invariance}

\author{Sophie Hermann}
\email{Sophie.Hermann@uni-bayreuth.de}
\affiliation{Theoretische Physik II, Physikalisches Institut, 
  Universit{\"a}t Bayreuth, D-95447 Bayreuth, Germany}
\author{Matthias Schmidt}
\email{Matthias.Schmidt@uni-bayreuth.de}
\affiliation{Theoretische Physik II, Physikalisches Institut, 
  Universit{\"a}t Bayreuth, D-95447 Bayreuth, Germany}

\date{14 September 2022}

\begin{abstract}
   The strength of fluctuations, as measured by their variance, is
   paramount in the quantitative description of a large class of
   physical systems, ranging from simple and complex liquids to active
   fluids and solids. Fluctuations originate from the irregular motion
   of thermal degrees of freedom and statistical mechanics facilitates
   their description.  Here we demonstrate that fluctuations are
   constrained by the inherent symmetries of the given system. For
   particle-based classical many-body systems, Noether invariance at
   second order in the symmetry parameter leads to exact sum rules.
   These identities interrelate the global force variance with the
   mean potential energy curvature. Noether invariance is
   restored by an exact balance between these distinct mechanisms. The
   sum rules provide a practical guide for assessing and constructing
   theories, for ensuring self-consistency in simulation work, and for
   providing a systematic pathway to the theoretical quantification of
   fluctuations.
\end{abstract}

\maketitle 

{\bf \noindent \large Introduction}\\
Applying Noether's theorem \cite{noether1918} to a physical problem
requires identifying and hence exploiting the fundamental symmetries
of the system under consideration.  Independent of whether such work
is performed in a Hamiltonian setting or on the basis of an action
functional, typically it is a conservation law that results from each
inherent symmetry of the system. The merits of the Noetherian strategy
have been demonstrated in a variety of contexts from classical
mechanics to field theory \cite{byers1998}. However, much of modern
condensed matter physics is focused on seemingly entirely different
physical behaviour, namely that of fluctuating, disordered, spatially
random, yet strongly interacting systems that possess a large number
of degrees of freedom.  Recent examples include active particles that
display freezing \cite{turci2021freezing} and wetting
\cite{turci2021wetting}, hydrophobicity rationalized as critical
drying \cite{coe2022}, the structure of two-dimensional colloidal
liquids \cite{thorneywork2018} and that of fluid interfaces
\cite{hoefling2015,parry2016}.

Relating the fluctuations that occur in complex systems to the
underlying symmetries has been investigated in a variety of
contexts. Such work addressed the symmetries in fluctuations far from
equilibrium \cite{hurtado2011}, isometric fluctuation relations
\cite{lacoste2014}, fluctuation relations for equilibrium states with
broken symmetry \cite{lacoste2015}, and fluctuation-response out of
equilibrium \cite{dechant2020}.  The fluctuation theorems of
stochastic thermodynamics provide a systematic setup to address such
questions \cite{seifert2012}.  Beyond its widespread use in
deterministic settings, Noether's theorem was formulated and used in a
stochastic context \cite{lezcano2018stochastic}, for Markov processes
\cite{baez2013markov}, for the quantification of the asymmetry
of quantum states \cite{marvian2014quantum}, for formulating entropy
as a Noether invariant \cite{sasa2016,sasa2020}, and for studying the
thermodynamical path integral and emergent symmetry \cite{sasa2019}.
Early work was carried out by Revzen~\cite{revzen1970} in the context of functional
integrals in statistical physics and a recent
perspective from an algebraic point of view was given by Baez
\cite{baez2020bottom}.

Noether's theorem has recently been suggested to be applicable in a
genuine statistical mechanical fashion
\cite{hermann2021noether,hermann2021noetherPopular,tschopp2022forceDFT}.
Based on translational and rotational symmetries the theorem allows to
derive exact identities (``sum rules'') with relative ease for
relevant many-body systems both in and out of equilibrium. The sum
rules set constraints on the global forces and torques in the system,
such as the vanishing of the global external force in equilibrium
\cite{hermann2021noether,baus1984} and of the global internal force
also in nonequilibrium \cite{hermann2021noether}. 

Here we demonstrate
that Noether's theorem allows to go beyond mere averages and
systematically address the strength of fluctuations, as measured by
the variance (auto-correlation).  We demonstrate that this variance is
balanced by the mean potential curvature, which hence restores the
Noether invariance. The structure emerges when going beyond the usual
linear expansion in the symmetry parameter.  The relevant objects to
be transformed are cornerstones of Statistical Mechanics, such as the
grand potential in its elementary form and the free energy density
functional. The invariances constrain both density fluctuations and
direct correlations, where the latter are generated from
  functional differentiation of the excess (over ideal gas) density
  functional.  \\

{\bf \noindent \large Results and Discussion}\\
{\bf \noindent External force variance.} We work in the grand ensemble and express the associated grand
potential in its elementary form \cite{hansen2013} as
\begin{align}
  & \Omega[V_{\rm ext}] =
  \label{EQOmegaElementary}
  \\&\quad
  -k_BT \ln \Tr 
  \exp\Big(-\beta\Big(H_{\rm int}
  +\sum_i V_{\rm ext}(\rv_i)-\mu N\Big)\Big),\notag
\end{align}
where $k_B$ indicates the Boltzmann constant, $T$ is absolute
temperature, and $\beta=1/(k_BT)$ is inverse temperature. The grand
ensemble ``trace'' is denoted by $\Tr = \sum_{N=0}^\infty 1/(N!h^{3N})
\int d\rv_1\ldots d\rv_N \int d\pv_1\ldots d\pv_N$, where $\rv_i$ is
the position and $\pv_i$ is the momentum of particle $i=1,\ldots,N$,
with $N$ being the total number of particles and $h$ the Planck
constant. The internal part of the Hamiltonian is $H_{\rm int} =
\sum_i\pv_i^2/(2m)+u(\rv_1,\ldots,\rv_N)$, where $m$ indicates the
particle mass, $u(\rv_1,\ldots,\rv_N)$ is the interparticle
interaction potential, and $V_\ext(\rv)$ is the external one-body
potential as a function of position $\rv$.  The thermodynamic
parameters are the chemical potential $\mu$ and temperature $T$.

Clearly, the value of the grand potential $\Omega[V_\ext]$ depends on
the function $V_{\rm ext}(\rv)$ and we have indicated this functional
dependence by the brackets. We consider a spatial displacement by a
constant vector $\eps$, applied to the entire system. The external
potential is hence modified according to $V_{\rm ext}(\rv)\to
V_\ext(\rv+\eps)$. This displacement leaves the kinetic energy
invariant (the momenta are unaffected) and it does not change the
interparticle potential $u(\rv_1,\ldots,\rv_N)$, as its dependence is
only on difference vectors $\rv_i-\rv_j$, which are unaffected by the
global displacement. 
Throughout we do not consider the dynamics of the shifting and rather
only compare statically the original with the displaced system, with
both being in equilibrium.  (Hermann and Schmidt~\cite{hermann2021noether} 
present dynamical Noether sum rules that arise from invariance of the power
functional \cite{schmidt2022rmp} at first order in a time-dependent
shifting protocol $\eps(t)$.)
The invariance with respect to the displacement can be explicitly seen
by transforming each position integral in the trace over phase space
as $\int d\rv_i=\int d(\rv_i-\eps)$. No boundary terms occur as the
integral is over $\mathbb R^3$; the effect of system walls is
explicitly contained in the form of $V_{\rm ext}(\rv)$. This
coordinate shift formally ``undoes'' the spatial system displacement
and it renders the form of the partition sum identical to that of the
original system. (See the work of Tschopp \textit{et al.}~\cite{tschopp2022forceDFT} 
for the generalization from homogeneous shifting to a position-dependent
strain operation.)  

The Taylor expansion of the grand potential of the displaced system
around the original system is
\begin{align}
  \Omega[V_\ext^\eps] &= \Omega[V_\ext]+
  \int d\rv \rho(\rv)\nabla V_\ext(\rv)\cdot\eps
  \notag\\
  &\quad+\frac{1}{2}
  \int d\rv \rho(\rv)
  \nabla\nabla V_\ext(\rv):\eps\eps
  \label{EQOmegaTaylorExpansionSecondOrder}
  \\
  &\quad -\frac{\beta}{2}\int d\rv d\rv'  
  H_2(\rv,\rv')
  \nabla V_\ext(\rv)\nabla' V_\ext(\rv'):\eps\eps,
  \notag
\end{align}
where we have truncated at second order in $\eps$ and have used the
shortcut notation $V_\ext^\eps(\rv)=V_\ext(\rv+\eps)$ for the
functional argument on the left hand side of
Eq.~\eqref{EQOmegaTaylorExpansionSecondOrder}. The colon indicates a
double tensor contraction and $\nabla V_\ext(\rv) \nabla'
V_\ext(\rv')$ is the dyadic product of the external force field with
itself. ($\nabla'$ denotes the derivative with respect to $\rv'$). The
occurrence of the one-body density profile $\rho(\rv)$ and of the
correlation function of density fluctuations $H_2(\rv,\rv')$ is due to
the functional identities $\rho(\rv)=\delta\Omega[V_\ext]/\delta
V_\ext(\rv)$ and $H_2(\rv,\rv')=-k_BT\delta^2\Omega[V_\ext]/\delta
V_\ext(\rv)\delta V_\ext(\rv')$
\cite{hansen2013,evans1979,evans1992,schmidt2022rmp}.

The Noetherian invariance against the displacement implies that the
value of the grand potential remains unchanged upon shifting, and
hence $\Omega[V_\ext^\eps]=\Omega[V_\ext]$
\cite{hermann2021noetherPopular}.  As a consequence, both the first
and the second order terms in the Taylor expansion
\eqref{EQOmegaTaylorExpansionSecondOrder} need to vanish identically,
and this holds irrespectively of the value of $\eps$; i.e.\ both
  the orientatation and the magnitude of $\eps$ can be arbitrary.
This yields, respectively, the first
\cite{hermann2021noether,baus1984} and second order 
  \cite{hirschfelder1960,haile1992} identities
\begin{align}
  -\int d\rv \rho(\rv)\nabla  V_\ext (&\rv) = 0,
  \label{EQNoetherFirstOrderVext}\\
  \int d\rv d\rv' H_2(\rv,\rv') & \nabla V_\ext(\rv)\nabla'V_\ext(\rv')=
  \notag \\
  &\quad 
  k_BT \int d\rv  \rho(\rv) \nabla\nabla V_\ext(\rv).
  \label{EQNoetherSecondOrderVext}
\end{align}
We can rewrite the sum rule \eqref{EQNoetherFirstOrderVext} in the
compact form $\langle\hat\Fv_\ext^\tot\rangle=0$, where we have
introduced the global external force operator $\hat\Fv_{\rm ext}^\tot
\equiv -\sum_i \nabla_i V_{\rm ext}(\rv_i)= -\int d\rv \sum_i
\delta(\rv-\rv_i) \nabla_i V_{\rm ext}(\rv_i)$.  The angular brackets
denote the equilibrium average $\langle\cdot\rangle = \Tr\Psi \cdot$,
where the grand ensemble distribution function is $\Psi={\rm
  e}^{-\beta(H-\mu N)}/\Xi$, with $H=H_{\rm int}+\sum_i V_\ext(\rv_i)$
and the grand partition sum is $\Xi=\Tr {\rm e}^{-\beta(H-\mu
  N)}$. Using these averages, and defining the density operator
$\hat\rho(\rv)=\sum_i\delta(\rv-\rv_i)$, where $\delta(\cdot)$ denotes
the Dirac distribution, allows to express the density profile as
$\rho(\rv)=\langle\sum_i\delta(\rv-\rv_i)\rangle$. The covariance of
the density operator is
$H_2(\rv,\rv')=\langle\hat\rho(\rv)\hat\rho(\rv')\rangle-\rho(\rv)\rho(\rv')$,
 which complements the above definition of $H_2(\rv,\rv')$ via the
  second functional derviative of the grand potential.

\begin{figure}
    \includegraphics[width=0.99\columnwidth,angle=0]{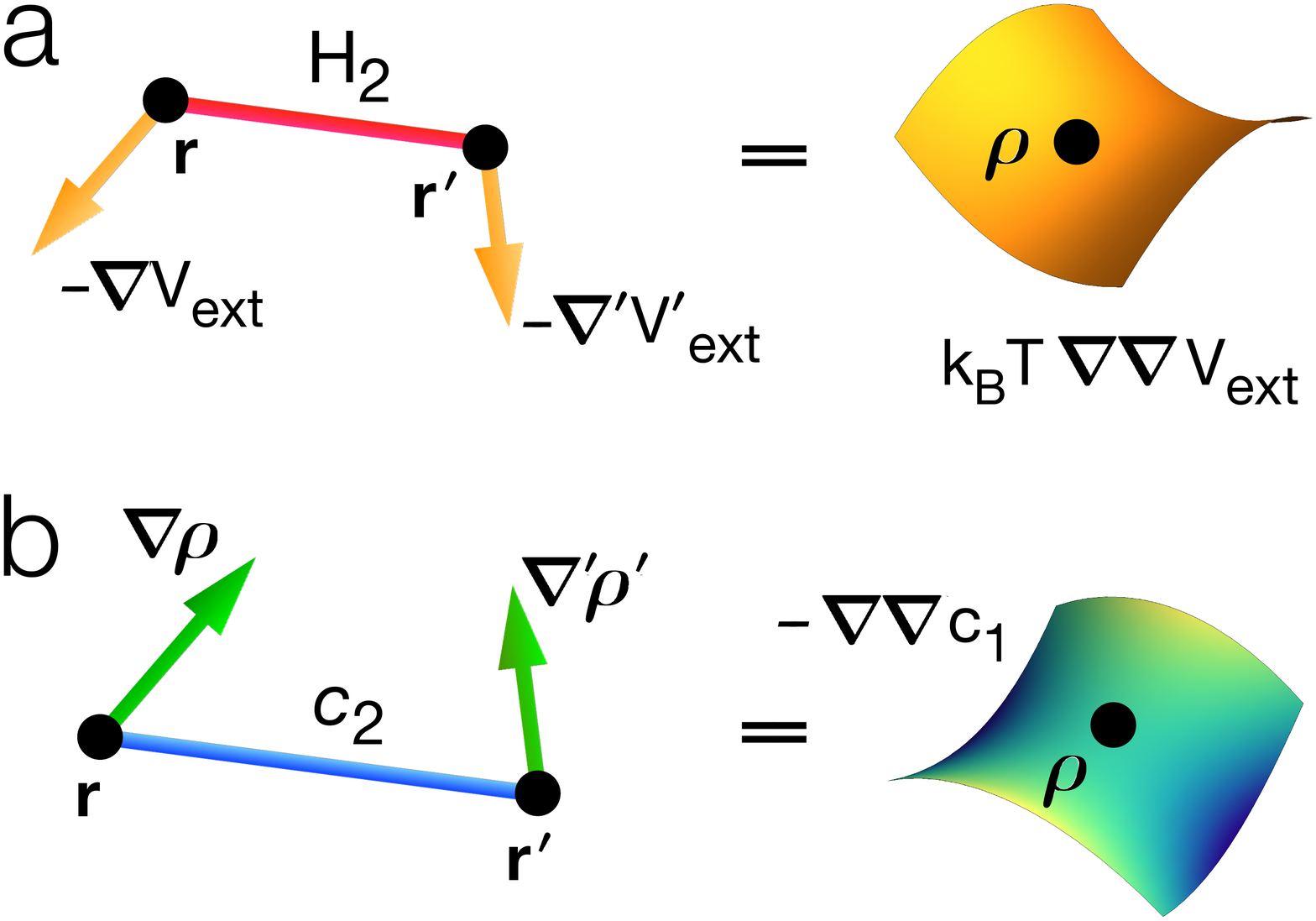}
  \caption{
  \textbf{Illustrations of the sum rules for the variance of
    fluctuations.}
    The sum rules arise from Noether invariance against
    spatial displacement. Shown are the different types of identical
    integrals. Thick dots indicate position variables that are
    integrated over. {\bf a} External sum rule,
    Eq.~\eqref{EQNoetherSecondOrderVext}, which relates the
    correlation function of density fluctuations $H_2(\rv,\rv')$ and
    the external force field $-\nabla V_{\rm ext}(\rv)$ with the
    product of the density profile $\rho(\rv)$ and the Hessian of the
    external potential $k_BT\nabla\nabla V_{\rm ext}(\rv).$  This 
    curvature is indicated by a schematic heat map. {\bf b}
    Internal sum rule, Eq.~\eqref{EQNoetherSecondOrderFexc}, where the
    density gradient at two different positions is bonded by the
    direct correlation function $c_2(\rv,\rv')$. This integral is
    identical to the integrated Hessian $-\nabla\nabla c_1(\rv)$ 
    (indicated by a schematic heat map)  weighted by the local 
    density $\rho(\rv)$.}
  \label{fig1}
\end{figure}

The second order sum rule~\eqref{EQNoetherSecondOrderVext} constrains
the variance of the external force operator on its left hand side:
$\langle \hat\Fv_\ext^\tot \hat\Fv_\ext^\tot\rangle - \langle
\hat\Fv_\ext^\tot\rangle\langle \hat\Fv_\ext^\tot\rangle = \langle
\hat\Fv_\ext^\tot \hat\Fv_\ext^\tot\rangle$; recall that the average
(first moment) of the external force vanishes, see
Eq.~\eqref{EQNoetherFirstOrderVext}.  The right hand side of
Eq.~\eqref{EQNoetherSecondOrderVext} balances the strength of these
force fluctuations by the mean curvature of the external potential
(multiplied by thermal energy $k_BT$), see Fig.~\ref{fig1}(a) for an
illustration of the structure of the integrals.

The curvature term can be re-written, upon integration by parts, as
$\int d\rv (-k_BT\nabla \rho(\rv)) \nabla V_\ext(\rv)$, which is the
integral of the local correlation of the ideal force density,
$-k_BT\nabla\rho(\rv)$, and the negative external force field $\nabla
V_\ext(\rv)$. (We assume setups with closed walls, where boundary
terms vanish.) The sum rule \eqref{EQNoetherSecondOrderVext} remains
valid if one replaces $H_2(\rv,\rv')$ by the two-body density
$\rho_2(\rv,\rv')=\langle\hat\rho(\rv)\hat\rho(\rv')\rangle$, due to
the vanishing of the external force \eqref{EQNoetherFirstOrderVext}.
 Explicitly, the alternative form of
  Eq.~\eqref{EQNoetherSecondOrderVext} that one obtains via this
  replacement is: $\int d\rv d\rv' \rho_2(\rv,\rv') \nabla
  V_\ext(\rv)\nabla' V_\ext(\rv')=k_BT \int d\rv \rho(\rv)
  \nabla\nabla V_\ext(\rv)$.

It is standard practice
\cite{hansen2013,evans1979,evans1992,schmidt2022rmp} to split off the
trivial density covariance of the ideal gas and define the total
correlation function $h(\rv,\rv')$ via the identity
$H_2(\rv,\rv')=\rho(\rv)\rho(\rv')h(\rv,\rv')
+\rho(\rv)\delta(\rv-\rv')$. Insertion of this relation into
Eq.~\eqref{EQNoetherSecondOrderVext} and then moving the term with the
delta function to the right hand side yields the following alternative
form of the second order Noether sum rule:
\begin{align}
 & \int d\rv d\rv' \rho\rho' h(\rv,\rv')
  \nabla V_\ext \nabla' V_\ext'=
  \notag\\
  &\qquad  \int d\rv 
  \big(k_BT\nabla\nabla V_{\rm ext}-(\nabla V_\ext)\nabla V_\ext
  \big)\rho,
  \label{EQNoetherWithLittleh} 
\end{align}
where we have left away the position arguments of the density
  profile and of the external potential for clarity and the prime
denotes dependence on $\rv'$. 
For the ideal gas $h(\rv,\rv')=0$ and hence the left hand side of
\eqref{EQNoetherWithLittleh} vanishes. That the right hand side then
also vanishes can be seen explicitly by inserting the generalized
barometric law \cite{hansen2013} $\rho(\rv)\propto
\exp(-\beta(V_\ext(\rv)-\mu))$ and either integrating by parts, or by
alternatively observing that $ -(k_BT)^2 \int d\rv
\nabla\nabla\rho(\rv)=0$ and inserting the barometric law therein.

The right hand side of \eqref{EQNoetherWithLittleh} makes explicit the
balancing of the external force variance with the mean potential
curvature, as given by its averaged Hessian.  For an interacting
(non-ideal) system, $h(\rv,\rv')$ is nonzero in general and the
  associated external force correlation contributions are accumulated
by the expression on the left hand side of
Eq.~\eqref{EQNoetherWithLittleh}.  For the special case of a harmonic
trap, as represented by the external potential $V_\ext(\rv)=\kappa
\rv^2/2$, with spring constant $\kappa$ and Hessian $\nabla\nabla
V_\ext(\rv)=\kappa \unity$, where $\unity$ denotes the unit matrix,
the mean curvature can be obtained explicitly. The first term on
  the right hand side of the sum rule \eqref{EQNoetherWithLittleh}
then simply becomes $k_BT \langle N \rangle \kappa\unity$ upon
  integration.  Notably, this result holds independently of the type
of interparticle interactions, although the latter affect
$h(\rv,\rv')$ as is present on the left hand side of
Eq.~\eqref{EQNoetherWithLittleh}.   The remaining (second) term on
  the right hand side of Eq.~\eqref{EQNoetherWithLittleh} turns into
  $-\kappa^2 \int d\rv \rho(\rv)\rv\rv$, where the integral is the
  matrix of second spatial moments of the density profile.  The
  alternative form $-\kappa^2 \langle \sum_i \rv_i\rv_i \rangle$ is
  obtained upon expressing the density profile as the average of
  $\hat\rho(\rv)$ and carrying out the integral over $\rv$.
  Collecting all terms and dividing by $\kappa^2$ we obtain the sum
  rule~\eqref{EQNoetherWithLittleh} for the case of an interacting
  system inside of a harmonic trap as: $\int d\rv d\rv'
  \rho(\rv)\rho(\rv')h(\rv,\rv')\rv \rv'= \int d\rv \rho(\rv) ( k_BT
  \kappa^{-1} \unity - \rv\rv)$.

{\bf \noindent Internal force variance.} In light of the external force fluctuations, one might wonder whether
the global interparticle force also fluctuates. The corresponding
operator is the sum of all interparticle forces: $\hat\Fv_{\rm
  int}^\tot \equiv -\sum_i \nabla_i u(\rv_1,\ldots,\rv_N)= -\int d\rv
\sum\delta(\rv-\rv_i)\nabla_i u(\rv_1,\ldots,\rv^N)$,  where the
  integrand in the later expression (including the minus sign) is the
  position-resolved force density operator
  \cite{schmidt2022rmp}. However, for each microstate $\hat\Fv_{\rm
  int}^\tot=0$, as can be seen e.g.\ via the translation invariance of
the interparticle potential \cite{hermann2021noether}, which
ultimately expresses Newton's third law {\it actio est reactio}. Hence
trivially the average vanishes, $\langle \hat\Fv_{\rm
  int}^\tot\rangle=0$, as do all higher moments, $\langle \hat\Fv_{\rm
  int}^\tot \hat\Fv_{\rm int}^\tot\rangle = 0$, as well as cross
correlations, $\langle \hat\Fv_{\rm int}^\tot \hat\Fv_{\rm
  ext}^\tot\rangle = 0$, etc. Thus the total internal force does
  not fluctuate. This holds beyond equilibrium, as the properties of
  the thermal average are not required in the argument. Identical
  reasoning can be applied to a nonequilibrium ensemble, where these
  identities hence continue to hold.

While these {\it probabilistic} correlators vanish, deeper inherent
structure can be revealed by addressing direct correlations, as
introduced by Ornstein and Zernike in 1914 in their treatment of
critical opalescence and to great benefit exploited in modern liquid
state theory \cite{hansen2013}. We use the framework of classical
density functional theory \cite{evans1979,evans1992,hansen2013}, where
the effect of the interparticle interactions is encapsulated in the
intrinsic Helmholtz excess free energy $F_{\rm exc}[\rho]$ as a
functional of the one-body density distribution $\rho(\rv)$.  As the
excess free energy functional solely depends on the interparticle
interactions, it necessarily is invariant against spatial
displacements. In technical analogy to the previous case of the
external force, we consider a displaced density profile
$\rho(\rv+\eps)$ and Taylor expand the excess free energy functional
up to second order in $\eps$ as follows:
\begin{align}
  \beta F_\exc[\rho^\eps] &= \beta F_\exc[\rho]
  -\int d\rv c_1(\rv) \nabla\rho(\rv)\cdot\eps\notag\\
  &\quad
  -\frac{1}{2}\int d\rv c_1(\rv)\nabla\nabla\rho(\rv):\eps\eps
  \label{EQTaylorFexc}
  \\
  &\quad
  -\frac{1}{2}\int d\rv d\rv' c_2(\rv,\rv')\nabla\rho(\rv)
  \nabla'\rho(\rv'):\eps\eps,
  \notag
\end{align}
where $\rho^\eps(\rv) = \rho(\rv+\eps)$ is again a shorthand.  The
one- and two-body direct correlation functions are given,
respectively, via the functional derivatives $c_1(\rv)=-\beta \delta
F_{\rm exc}[\rho]/\delta\rho(\rv)$ and $c_2(\rv,\rv')=-\beta\delta^2
F_{\rm exc}[\rho]/\delta\rho(\rv)\delta\rho(\rv')$.  Noether
invariance demands that $F_{\rm exc}[\rho^\eps]=F_{\rm exc}[\rho]$ and
hence both the linear and the quadratic contributions in the Taylor
expansion \eqref{EQTaylorFexc} need to vanish, irrespective of the
value of $\eps$. This yields, respectively:
\begin{align}
  \int d\rv c_1(\rv)\nabla\rho(\rv) &= 0,
  \label{EQNoetherFirstOrderFexc}\\
  \int d\rv d\rv' c_2(\rv,\rv')\nabla\rho(\rv)\nabla'\rho(\rv')
  &= -\int d\rv\rho(\rv) \nabla\nabla c_1(\rv),
  \label{EQNoetherSecondOrderFexc}
\end{align}
where we have integrated by parts on the right hand side of
\eqref{EQNoetherSecondOrderFexc}.
The first order sum rule \eqref{EQNoetherFirstOrderFexc} expresses the
vanishing of the global internal force $\langle\hat\Fv_{\rm
  int}^\tot\rangle=0$ \cite{hermann2021noether}. This can be seen by
integrating by parts, which yields the integrand in the form
$-\rho(\rv)\nabla c_1(\rv)$, which is the internal force density
scaled by $-k_BT$.  In formal analogy to the probabilistic variance in
Eq.~\eqref{EQNoetherSecondOrderVext}, the second order sum rule
\eqref{EQNoetherSecondOrderFexc} could be viewed as relating the
``direct variance'' of the density gradient (left hand side) to the
mean gradient of the internal one-body force field in units of $k_BT$
(right hand side), which, equivalently, is the Hessian of the local
intrinsic chemical potential $-k_BT c_1(\rv)$, see Fig.~\ref{fig1}(b).

 As a conceptual point concerning the derivations of
  Eqs.~\eqref{EQNoetherFirstOrderFexc} and
  \eqref{EQNoetherSecondOrderFexc}, we point out that the excess free
  energy density functional $F_{\rm exc}[\rho]$ is an intrinsic
  quantity, which does not explicitly depend on the external potential
  $V_{\rm ext}(\rv)$. Hence there is no need to explicitly take into
  account a corresponding shift of $V_{\rm ext}(\rv)$. This is true
  despite the fact that in an equilibrium situation one would consider
  the external potential (and the correspondingly generated external
  force field) as the physical reason for the (inhomogeneous) density
  profile to be stable. Both one-body fields are connected via the
  (Euler-Lagrange) minimization equation of density functional theory
  \cite{evans1979,evans1992,hansen2013}:
  $k_BT\ln\rho(\rv)=k_BTc_1(\rv)- V_{\rm ext}(\rv)+\mu$, where we have
  set the thermal de Broglie wavelength to unity.  For given density
  profile, we can hence trivially obtain the corresponding external
  potential as $V_{\rm ext}(\rv)=-k_BT\ln\rho(\rv)+k_BTc_1(\rv)+\mu$,
  which makes the fundamental Mermin-Evans
  \cite{evans1979,hansen2013,evans1992,schmidt2022rmp} map
  $\rho(\rv)\to V_{\rm ext}(\rv)$ explicit.

As a consistency check, the second order sum rules
\eqref{EQNoetherSecondOrderVext} and \eqref{EQNoetherSecondOrderFexc}
can alternatively be derived from the hyper virial theorm
  \cite{hirschfelder1960,haile1992} or from spatially resolved
correlation identities \cite{hermann2021noether,baus1984}. Following
the latter route, one starts with $\int
d\rv'H_2(\rv,\rv')\nabla'V_\ext(\rv')= -k_BT\nabla\rho(\rv)$ and $\int
d\rv'c_2(\rv,\rv')\nabla'\rho(\rv')=\nabla c_1(\rv)$, respectively.
The derivation the requires the choice of a suitable field as a
multiplier ($\nabla V_\ext(\rv)$ and $\nabla\rho(\rv)$, respectively),
spatial integration over the free position variable, and subsequent
integration by parts.  However, this strategy i) requires the correct
choice for multiplication to be made, and ii) it does not allow to
identify the Noether invariance as the underlying reason for the
validity.  In contrast, the Nother route is constructive and it allows
to trace spatial invariance as the fundamental physical reason for the
respective identity to hold.

{\bf \noindent Thermal diffusion force variance.} Similar to the treatment of the excess free energy functional, one can
shift and expand the ideal free energy functional $F_{\rm
  id}[\rho]=k_BT\int d\rv \rho(\rv)(\ln\rho(\rv)-1)$.  Exploiting the
translational invariance at first order leads to vanishing of the
total diffusive force: $-k_BT \int d\rv \nabla \rho(\rv) = 0$, and at
second order: $\int d\rv \rho(\rv)^{-1}(\nabla \rho(\rv))
\nabla\rho(\rv) = -\int d\rv \rho(\rv) \nabla \nabla \ln
\rho(\rv)$. These ideal identities can be straightforwardly verified
via integration by parts (boundary contributions vanish) and they
complement the excess results \eqref{EQNoetherFirstOrderFexc} and
\eqref{EQNoetherSecondOrderFexc}.

{\bf \noindent Outlook.} While we have restricted ourselves throughout to translations in
equilibrium, the variance considerations apply analogously for
rotational invariance \cite{hermann2021noether} and to the dynamics,
where invariance of the power functional forms the basis
\cite{hermann2021noether,schmidt2022rmp}. In future work it it would
be highly interesting to explore connections of our results to
statistical thermodynamics \cite{seifert2012}, to the study of liquids
under shear \cite{asheichyk2021}, to the large fluctuation functional
\cite{jack2015}, as well as to recent progress in systematically
incorporating two-body correlations into classical density functional
theory \cite{tschopp2020,tschopp2021}.  Investigating the implications
of our variance results for Levy-noise \cite{yuvan2022} is
interesting. As the displacement vector $\eps$ is arbitrary both
  in its orientation and its magnitude our reasoning does not stop at
  second order in the Taylor expansion, see
  Eqs.~\eqref{EQOmegaTaylorExpansionSecondOrder} and
  \eqref{EQTaylorFexc}. Assuming that the power series exists, the
  invariance against the displacement rather implies that each order
  vanishes individually, which gives rise to a hierarchy of
  correlation identities of third, fourth, {\it etc.} moments that are
  interrelated with third, fourth, {\it etc.} derivatives of the
  external potential (when starting from $\Omega[V_{\rm ext}]$) or the
  one-body direct correlation function (when starting from the excess
  free energy density functional $F_{\rm exc}[\rho]$). 

 Future use of the sum rules can be manifold, ranging from the
construction and testing of new theories, such as approximate free
energy functionals within the classical density functional framework,
to validation of simulation data (to ascertain both correct
  implementation and sufficient equilibration and sampling) and
numerical theoretical results. To give a concrete example, in
  systems like the confined hard sphere liquid considered by Tschopp 
  \textit{et al.}~\cite{tschopp2022forceDFT} on the basis of fundamental measure
  theory, one could apply and test the sum rule
  \eqref{EQNoetherWithLittleh} explicitly, as the inhomogeneous total
  pair correlation function $h(\rv,\rv')$ is directly accessible in
  the therein proposed force-DFT approach.\\

{\bf \noindent \large Data availability}\\ Data sharing is not
  applicable to this study as no datasets were generated or analyzed
  during the current study.\\

{\bf \noindent \large Acknowledgments}\\
Open Access funding enabled and organized by Projekt DEAL. 
We thank Daniel de las Heras, Thomas Fischer, and Gerhard Jung
 for useful discussions.\\

{\bf \noindent \large Author contributions}\\
S.H. and M.S. have jointly carried out the work and written the paper.\\


{\bf \noindent \large Competing interests}\\
The authors declare no competing interests.


\begin{thebibliography}{31}


\bibitem{noether1918} E. Noether, {\it Invariante Variationsprobleme,}
  \href{https://gdz.sub.uni-goettingen.de/download/pdf/PPN252457811_1918/LOG_0022.pdf}
       {Nachr. d. K\"onig. Gesellsch. d. Wiss. zu G\"ottingen,
         Math.-Phys. Klasse, 235 (1918).}
  English translation by M. A. Tavel: {\it Invariant variation
  problems,} Transp. Theo. Stat.  Phys. {\bf 1}, 186 (1971); for a
  version in modern typesetting see: Frank Y. Wang,
  \href{http://arxiv.org/abs/physics/0503066v3}{arXiv:physics/0503066v3}
  (2018).

\bibitem{byers1998} 
  N. Byers,
  {\it E.\ Noether's discovery of the deep connection between
  symmetries and conservation laws,}
  \href{https://arxiv.org/abs/physics/9807044}
       {arXiv:physics/9807044 (1998)}.

\bibitem{turci2021freezing}
  F. Turci and N. B. Wilding,
  {\it Phase separation and multibody effects in three-dimensional active Brownian
    particles,}
  \href{https://doi.org/10.1103/PhysRevLett.126.038002}
       {Phys. Rev. Lett. {\bf 126}, 038002 (2021).}

\bibitem{turci2021wetting}       
  F. Turci and N. B. Wilding,
  {\it Wetting transition of active Brownian particles on a thin membrane,}
  \href{https://doi.org/10.1103/PhysRevLett.127.238002}
       {Phys. Rev. Lett. {\bf 127}, 238002 (2021).}

\bibitem{coe2022}
  M. K. Coe, R. Evans, and N. B. Wilding,
  {\it Density depletion and enhanced fluctuations in water near
    hydrophobic solutes: identifying the underlying physics,}
  \href{https://doi.org/10.1103/PhysRevLett.128.045501}
       {Phys. Rev. Lett. {\bf 128}, 045501 (2022).}

\bibitem{thorneywork2018}
  A. L. Thorneywork, S. K. Schnyder, D. G. A. L. Aarts, J. Horbach, R. Roth, 
  and R. P. A. Dullens,
  {\it Structure factors in a two-dimensional binary colloidal hard sphere system,}
  \href{https://doi.org/10.1080/00268976.2018.1492745}
       {Mol. Phys. {\bf 116}, 3245 (2018).}

\bibitem{hoefling2015}
  F. H\"ofling and S. Dietrich,
  {\it Enhanced wavelength-dependent surface tension of liquid-vapour interfaces,}
  \href{https://doi.org/10.1209/0295-5075/109/46002}
       {Europhys. Lett. {\bf 109}, 46002 (2015).}

\bibitem{parry2016}
  A. O. Parry, C Rasc\'on, and R. Evans,
  {\it The local structure factor near an interface;
    beyond extended capillary-wave models,}
  \href{http://dx.doi.org/10.1088/0953-8984/28/24/244013}
       {J. Phys.: Condens. Matter {\bf 28}, 244013 (2016).}

\bibitem{hurtado2011}
  P. I. Hurtado, C. P\'erez-Espigares, J. J. del Pozo, and P. L. Garrido,
  {\it Symmetries in fluctuations far from equilibrium},
  \href{https://doi.org/10.1073/pnas.1013209108}
       {Proc. Natl. Acad. Sci. {\bf 108}, 7704 (2011)}.

\bibitem{lacoste2014}
  D. Lacoste and P. Gaspard,
  {\it Isometric fluctuation relations for equilibrium states with broken symmetry,}
  \href{http://dx.doi.org/10.1103/PhysRevLett.113.240602}
       {Phys. Rev. Lett.  {\bf 113}, 240602  (2014)}.

\bibitem{lacoste2015}
  D. Lacoste and P. Gaspard,
  {\it Fluctuation relations for equilibrium states with broken 
    discrete or continuous symmetries,}
  \href{https://doi.org/10.1088/1742-5468/2015/11/P11018}
  {J. Stat. Mech.  {\bf 2015}, P11018 (2015)}.


\bibitem{dechant2020}
  A. Dechant and S. Sasa,
  {\it Fluctuation-response inequality out of equilibrium},
  \href{https://doi.org/10.1073/pnas.1918386117}
       {Proc. Natl. Acad. Sci. {\bf 117}, 6430 (2020)}.

\bibitem{seifert2012}
  U. Seifert,
  {\it Stochastic thermodynamics, fluctuation theorems and molecular machines,}
  \href{https://doi.org/10.1088/0034-4885/75/12/126001}
       {Rep. Prog. Phys. {\bf 75}, 126001 (2012).}

\bibitem{lezcano2018stochastic}
  A. G. Lezcano and A. C. M. de Oca, 
  {\it A stochastic version of the Noether theorem,}
  \href{https://doi.org/10.1007/s10701-018-0174-z}
       {Found. Phys. \textbf{48}, 726 (2018).}

\bibitem{baez2013markov}
  J. C. Baez and B. Fong, 
  {\it A Noether theorem for Markov processes,}
  \href{http://dx.doi.org/10.1063/1.4773921}
       {J. Math. Phys. \textbf{54}, 013301 (2013).}

\bibitem{marvian2014quantum}
  I. Marvian and R. W. Spekkens, 
  {\it Extending Noether’s theorem by quantifying the asymmetry 
    of quantum states,}
  \href{https://doi.org/10.1038/ncomms4821}
       {Nat. Commun. \textbf{5}, 3821 (2014).}

\bibitem{sasa2016}
   S. Sasa and Y. Yokokura,
   {\it Thermodynamic entropy as a Noether invariant,}
   \href{http://dx.doi.org/10.1103/PhysRevLett.116.140601}
   {Phys. Rev. Lett. {\bf 116}, 140601 (2016).}
   
\bibitem{sasa2020}
  Y. Minami and S. Sasa, 
  {\it Thermodynamic entropy as a Noether invariant in a Langevin equation,}
  \href{https://doi.org/10.1088/1742-5468/ab5b8b}
       {J. Stat. Mech. {\bf 2020} 013213 (2020).}

\bibitem{sasa2019} 
  S. Sasa, S. Sugiura, and Y. Yokokura, 
  {\it Thermodynamical path integral and emergent symmetry,}
  \href{https://doi.org/10.1103/PhysRevE.99.022109}
       {Phys. Rev. E \textbf{99}, 022109 (2019).}

\bibitem{revzen1970}
  M. Revzen,
  {\it Functional integrals in statistical physics,}
  \href{https://doi.org/10.1119/1.1976414}
       {Am. J. Phys. {\bf 38}, 611 (1970).}

\bibitem{baez2020bottom}
  J. C. Baez,
  {\it Getting to the Bottom of Noether's Theorem,}
  \href{https://arxiv.org/abs/2006.14741}
       {arXiv:2006.14741} (2022).

\bibitem{hermann2021noether}
  S. Hermann and M. Schmidt, 
  {\it Noether's Theorem in Statistical Mechanics,}
  \href{https://doi.org/10.1038/s42005-021-00669-2}
       {Commun. Phys.  {\bf  4}, 176 (2021).}

\bibitem{hermann2021noetherPopular}
  S. Hermann and M. Schmidt, 
  {\it Why Noether's Theorem applies to Statistical Mechanics,}
  \href{https://doi.org/10.1088/1361-648X/ac5b47}
       {J. Phys.: Condens. Matter
          {\bf 34}, 213001 (2022) (invited Topical Review).}

\bibitem{tschopp2022forceDFT}
  S. M. Tschopp, F. Samm\"uller, S. Hermann, M. Schmidt, and J. M. Brader,
  {\it Force density functional theory for fluids in- and 
    out-of-equilibrium,}
  \href{https://doi.org/10.1103/PhysRevE.106.014115}
        Phys. Rev. E {\bf 106}, 014115 (2022).
       
\bibitem{baus1984}
  M. Baus, 
  {\it Broken symmetry and invariance properties of classical fluids,}
  \href{https://doi.org/10.1080/00268978400100161}
  {Mol. Phys. {\bf 51}, 211 (1984).}

\bibitem{hansen2013} 
  J.-P. Hansen and I.~R. McDonald, {\it Theory of
  Simple Liquids}, 4th ed.\  (Academic Press, London, 2013). 

\bibitem{evans1979} R. Evans,
  {\it The nature of the liquid-vapour
  interface and other topics in the statistical mechanics of
  non-uniform, classical fluids,} 
  \href{https://doi.org/10.1080/00018737900101365}
  {Adv. Phys. {\bf 28}, 143 (1979).}

\bibitem{evans1992}
  R. Evans,
  {\it Density functionals in the theory nonuniform fluids,}
  in: Fundamentals of Inhomogeneous Fluids,
  edited by D. Henderson (Dekker, New York, 1992).

\bibitem{schmidt2022rmp}
  M. Schmidt,
  {\it Power functional theory for many-body dynamics},
  \href{https://doi.org/10.1103/RevModPhys.94.015007}
  {Rev. Mod. Phys. {\bf 94}, 015007 (2022).}

\bibitem{hirschfelder1960}
  J. O. Hirschfelder,
  {\it Classical and Quantum Mechanical Hypervirial Theorems,}
  \href{https://doi.org/10.1063/1.1731427}
       {J. Chem. Phys. {\bf 33}, 1462 (1960).}

\bibitem{haile1992}
  J. M. Haile,
  {\it Molecular Dynamics Simulation: Elementary Methods},
  (Wiley, New York, 1992).


\bibitem{asheichyk2021}
  K. Asheichyk, M. Fuchs, and M. Kr\"uger,
  {\it Brownian systems perturbed by mild shear: comparing response relations,}
  \href{https://doi.org/10.1088/1361-648X/ac0c3c}
  {J. Phys.: Condens. Matter {\bf 33}, 405101 (2021). }

\bibitem{jack2015}
  R. L. Jack and P. Sollich,
  {\it Effective interactions and large deviations in stochastic processes,}
  \href{http://dx.doi.org/10.1140/epjst/e2015-02416-9}
  {Eur. Phys. J. Special Topics {\bf 224}, 2351 (2015).}

\bibitem{tschopp2020}
  S. M. Tschopp, H.~D. Vuijk, A. Sharma, and J.~M. Brader,
  {\it Mean-field theory of inhomogeneous fluids,}
  \href{https://doi.org/10.1103/PhysRevE.102.042140}
  {Phys. Rev. E {\bf 102}, 042140 (2020).}

\bibitem{tschopp2021}
  S. M. Tschopp and J.~M. Brader,
  {\it Fundamental measure theory of inhomogeneous
    two-body correlation functions,}
  \href{https://doi.org/10.1103/PhysRevE.103.042103}
  {Phys. Rev. E {\bf 103}, 042103 (2021)}.

\bibitem{yuvan2022}
  S. Yuvan and M. Bier,
  {\it Accumulation of particles and formation of a dissipative structure in
    a nonequilibrium bath,}
  \href{https://doi.org/10.3390/e24020189}
  {Entropy {\bf 24}, 189 (2022).}\\

\end{thebibliography}
\end{document}